\documentclass[aip,jcp,reprint]{revtex4-1}

\usepackage[utf8]{inputenc}
\usepackage[fleqn]{amsmath}
\usepackage{amsfonts}
\usepackage{amssymb}
\usepackage{amsthm}
\usepackage{amsopn}
\usepackage{upgreek}
\usepackage{bm}
\usepackage{epsfig}
\usepackage{relsize}
\usepackage{hyperref}

\newcommand{\Zid}{Z_{\text{id}}}
\newcommand{\Zin}{Z_{\text{in}}}
\newcommand{\Fin}{F_{\text{in}}}
\newcommand{\qc}{q_{\text{c}}}
\newcommand{\qd}{q_{\text{d}}}
\newcommand{\be}{\beta\epsilon}

\newcommand{\bte}{\beta\tilde\epsilon}

\newcommand{\barzEV}{\bar{z}_{E,V}}

\newcommand{\barBEV}{\bar{B}_{E,V}}
\newcommand{\bteEV}{\bte_{E,V}}
\newcommand{\av}{\text{seq}}

\newcommand{\kB}{k_{\mathrm{B}}}

\begin{document}

\renewcommand{\eqref}[1]{Eq.~(\ref{#1})}
\newcommand{\eqsref}[2]{Eqs.~(\ref{#1}-\ref{#2})}
\newcommand{\figref}[1]{Figure~\ref{#1}}
\newcommand{\figsref}[1]{Figures~\ref{#1}}
\newcommand{\secref}[1]{Sec.~\ref{#1}}
\newcommand{\secsref}[1]{Secs.~\ref{#1}}
\newcommand{\appref}[1]{Appendix~\ref{#1}}
\newcommand{\refcite}[1]{Ref.~\onlinecite{#1}}
\newcommand{\refscite}[1]{Refs.~\onlinecite{#1}}

\newcommand{\addresscambridge}{Department of Chemistry, University of Cambridge, Lensfield Road, Cambridge CB2 1EW, United Kingdom}

\title{Rational design of self-assembly pathways for complex multicomponent structures}
\author{William M.~Jacobs}
\affiliation{\addresscambridge}
\author{Aleks~Reinhardt}
\affiliation{\addresscambridge}
\author{Daan~Frenkel}
\affiliation{\addresscambridge}
\date{\today}

\begin{abstract}
The field of complex self-assembly is moving toward the design of multi-particle structures consisting of thousands of distinct building blocks.
To exploit the potential benefits of structures with such `addressable complexity,' we need to understand the factors that optimize the yield and the kinetics of self-assembly.
Here we use a simple theoretical method to explain the key features responsible for the unexpected success of DNA-brick experiments, which are currently the only demonstration of reliable self-assembly with such a large number of components.
Simulations confirm that our theory accurately predicts the narrow temperature window in which error-free assembly can occur.
Even more strikingly, our theory predicts that correct assembly of the complete structure may require a time-dependent experimental protocol.
Furthermore, we predict that low coordination numbers result in non-classical nucleation behavior, which we find to be essential for achieving optimal nucleation kinetics under mild growth conditions.
We also show that, rather surprisingly, the use of heterogeneous bond energies improves the nucleation kinetics and in fact appears to be necessary for assembling certain intricate three-dimensional structures.
This observation makes it possible to sculpt nucleation pathways by tuning the distribution of interaction strengths.
These insights not only suggest how to improve the design of structures based on DNA bricks, but also point the way toward the creation of a much wider class of chemical or colloidal structures with addressable complexity.
\end{abstract}

\maketitle

Recent experiments with short pieces of single-stranded DNA\cite{wei2012complex,ke2012three} have shown that it is possible to assemble well-defined molecular superstructures from a single solution with more than merely a handful of distinct building blocks.
These experiments use complementary DNA sequences to encode an addressable structure\cite{rothemund2000program} in which each distinct single-stranded `brick' belongs in a specific location within the target assembly.
A remarkable feature of these experiments is that even without careful control of the subunit stoichiometry or optimization of the DNA sequences, a large number of two- and three-dimensional designed structures with thousands of subunits assemble reliably.\cite{wei2012complex,ke2012three,zhang2013self,wei2013design}
The success of this approach is astounding given the many ways in which the assembly of an addressable structure could potentially go wrong.\cite{rothemund2012nanotechnology,gothelf2012lego,kim2008probing}

Any attempt to optimize the assembly yield or to create even more complex structures should be based on a better understanding of the mechanism by which DNA bricks manage to self-assemble robustly.
The existence of a sizable nucleation barrier, as originally proposed in \refscite{wei2012complex} and \onlinecite{ke2012three}, would remedy two possible sources of error that were previously thought to limit the successful assembly of multicomponent nanostructures: the depletion of free monomers and the uncontrolled aggregation of partially formed structures.
Slowing the rate of nucleation would suppress competition among multiple nucleation sites for available monomers and give the complete structure a chance to assemble before encountering other partial structures.
Recent simulations of a simplified model of a three-dimensional addressable structure have provided evidence of a free-energy barrier for nucleation,\cite{reinhardt2014numerical} suggesting that the ability to control this barrier should enable the assembly of a wide range of complex nanostructures.
We therefore need to be able to predict how such a barrier depends on the design of the target structure and on the choice of DNA sequences.
Until now, however, there have been no reliable techniques to predict the existence, let alone the magnitude, of a nucleation barrier for self-assembly in a mixture of complementary DNA~bricks.

Here we show that the assembly of three-dimensional DNA-brick nanostructures is indeed a nucleated process, but only in a narrow range of temperatures.
The nucleation barrier in these systems is determined entirely by the topology of the designed interactions that stabilize the target structure.
Controllable nucleation is therefore a general feature of addressable structures that can be tuned through the rational choice of designed interactions.
We find that the reliable self-assembly of three-dimensional DNA bricks is a direct consequence of their unusual nucleation behavior, which is not accounted for by existing theories that work for classical examples of self-assembly, such as crystal nucleation.
We are thus able to provide a rational basis for the rather unconventional protocol used in the recent DNA-brick experiments by showing that they exploit a narrow window of opportunity where robust multicomponent self-assembly can take place.

\begin{figure*}[t!]
  \centering
  \includegraphics[width=17.5cm]{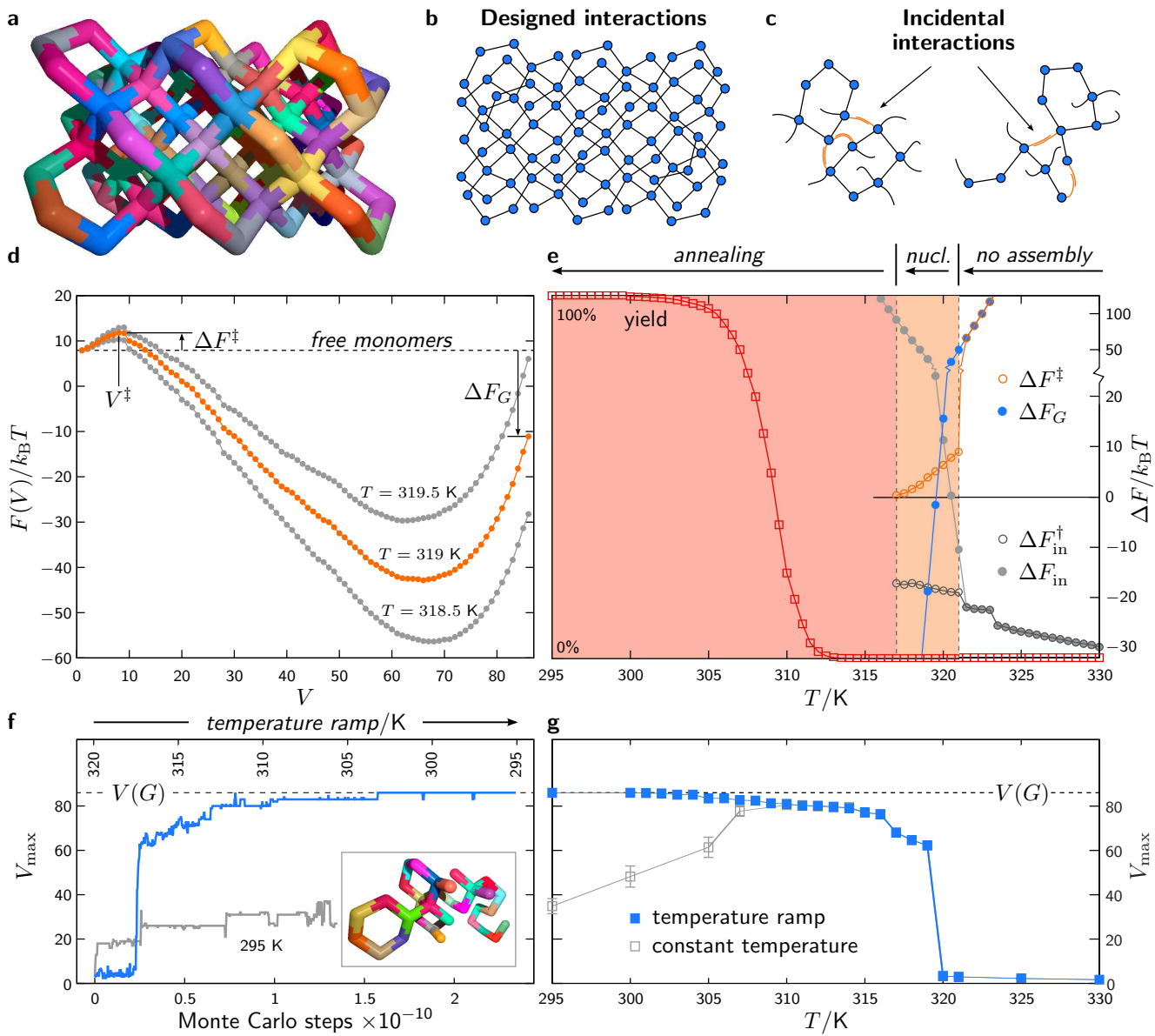}
  \caption{Controlled nucleation is essential for robust self-assembly.  (a)~An example 86-strand DNA-brick structure and (b)~its associated connectivity graph.  (c)~Incidental interactions between dangling ends, shown in orange, lead to incorrect associations between fragments.  (d)~Free energies of clusters of $V$ subunits with randomly chosen DNA sequences in units of $\kB T$, where $\kB$ is the Boltzmann constant and $T$ is the absolute temperature.  The nucleation barrier, $\Delta F^\ddagger$, and the target structure stability, $\Delta F_G$, are strongly temperature-dependent.  (e)~The equilibrium yield with exclusively designed interactions and the nucleation barrier as a function of temperature.  Also shown are the target structure stability and the free-energy difference between all off-pathway and on-pathway intermediates, in the presence, $\Delta \Fin^\dagger$, and absence, $\Delta \Fin^{\protect\phantom{\dagger}}$, of a nucleation barrier.  The nucleation window is shown in orange.  (f)~Representative lattice Monte Carlo simulation trajectories with (blue) and without (gray) a temperature ramp; a typical malformed structure is shown in the inset.  (g)~The size of the largest correctly bonded stable cluster in lattice Monte Carlo simulations using a temperature ramp (blue) and in constant-temperature simulations initialized from free monomers in solution (gray).}
  \label{fig:ramp_example}
\end{figure*}

\section*{Structure connectivity determines assembly}

In constructing a model for the self-assembly of addressable structures, we note that the designed interactions should be much stronger than any attractive interactions between subunits that are not adjacent in a correctly assembled structure.
The designed interactions that stabilize the target structure can be described by a connectivity graph, $G$, in which each vertex represents a distinct subunit and each edge indicates a correct bond.
This graph allows us to describe the connectivity of the structure independently of the geometry and spatial organization of the building blocks.
For structures constructed from DNA bricks, the edges of $G$ indicate the hybridization of DNA strands with complementary sequences that are adjacent in the target structure.
An example three-dimensional DNA-brick structure is shown along with its connectivity graph in \figsref{fig:ramp_example}a and \ref{fig:ramp_example}b.

In an ideal solution with exclusively designed interactions, the subunits assemble into clusters in which all allowed bonds are encoded in the connectivity graph of the target structure.
In order to compute the free-energy difference between a particular cluster size and the unbonded single-stranded bricks, we must consider all the ways in which a correctly bonded cluster with a given number of monomers can be assembled.
These `fragments' of the target structure correspond to connected subgraphs of the connectivity graph.
In a dilute solution with strong designed interactions, the numbers of edges and vertices are the primary factors determining the stability of a particular fragment.
We therefore identify all of the possible on-pathway assembly intermediates by grouping fragments into sets with the same number of edges and vertices and counting the total number of fragments in each set.\cite{jacobs2015theoretical}

This theoretical approach is powerful because it can predict the free-energy landscape as a function of the degree of assembly between the monomers and the target structure.
Furthermore, the predicted landscape captures the precise topology of the target structure, which is essential for understanding the assembly of addressable, finite-sized structures.
In the case of DNA-brick structures, we can assign DNA hybridization free energies to the edges of the target connectivity graph in order to determine the temperature dependence of the free-energy landscape; for example, \figref{fig:ramp_example}d shows the free-energy profile of the 86-strand DNA-brick structure with random DNA sequences at three temperatures.
Our theoretical approach allows us to calculate the nucleation barrier, $\Delta F^\ddagger$, by examining the free energies of clusters corresponding to fragments with exactly $V$ vertices.
The critical number of strands required for nucleation is $V^\ddagger$: transient clusters with fewer than $V^\ddagger$ strands are more likely to dissociate than to continue incorporating additional strands.
The presence of a substantial nucleation barrier therefore inhibits the proliferation of large, partially assembled fragments that stick together to form non-target aggregates.

\section*{Assembly requires a time-dependent protocol}

Over a significant range of temperatures, we find that the free-energy profiles of DNA-brick structures exhibit both a nucleation barrier and a thermodynamically stable intermediate structure.
The nucleation barrier is associated with the minimum number of subunits that must be assembled in order to complete one or more \textit{cycles}, i.e.~closed loops of stabilizing bonds in a fragment.
For example, the critical number of monomers in the example structure at 319~K, ${V^\ddagger = 8}$, is one fewer than the nine subunits required to form a bicyclic fragment of the target structure.
Under the conditions where nucleation is rate controlling, the minimum free-energy structure is \textit{not} the complete 86-particle target structure, but rather a structure with only  ${V \simeq 65}$ particles.
This incomplete structure is favored by entropy, since it can be realized in many more ways than the unique target structure.
Hence, the temperature where nucleation is rate controlling is higher than the temperature where the target structure is the most stable cluster.
The existence of thermodynamically stable intermediates is a typical feature of DNA-brick structures  and of complex addressable, finite-sized structures in general.\footnote{We note that thermodynamically stable partial structures have also been observed in previously reported simulations\cite{reinhardt2014numerical} of DNA-brick structures consisting of approximately 1,000 distinct subunits.}

This behavior is not compatible with classical nucleation theory (CNT), which predicts that, beyond the nucleation barrier, large clusters are always more stable than smaller clusters.
As a consequence, in `classical' nucleation scenarios such as crystallization, there is a sharp boundary in temperature and concentration at which the largest-possible ordered structure, rather than the monomeric state, becomes thermodynamically stable.
Typically, a simple fluid must be supersaturated well beyond this boundary in order to reduce the nucleation barrier, which arises due to the competition between the free-energy penalty of forming a solid--liquid interface and the increased stability due to the growth of an ordered structure.\cite{gibbs1878equilibrium,oxtoby1992homogeneous,sear2007nucleation}
Yet in the case of addressable self-assembly, and DNA bricks in particular, a nucleation barrier for the formation of a stable partial structure may exist even when the target structure is unstable relative to the free monomers.

An experiment to assemble such a structure requires a {\em protocol}: first nucleation at a relatively high temperature, and then further cooling to complete the formation of the target structure.\footnote{DNA-brick structures have also been assembled at constant temperature by changing the solution conditions during the course of the experiment.\cite{zhang2013self}  For simplicity, we assume that the experimental control parameter is the temperature, but clearly other methods of altering the DNA hybridization free energies during the assembly process may also be suitable.}
This behavior can be seen in \figref{fig:ramp_example}e, where we identify a narrow temperature window in which there is a significant yet surmountable nucleation barrier.
Unlike CNT, the nucleation barrier does not diverge as the temperature is increased.
Instead, there is a well-defined temperature above which all clusters have a higher free energy than the free monomers.
As the temperature is lowered further, the nucleation barrier disappears entirely before the equilibrium yield, defined as the fraction of all clusters that are correctly assembled as the complete target structure, increases measurably above zero.
The equilibrium yield tends to 100\% at low temperatures, since we have thus far assumed that only designed interactions are possible.
Therefore, because of the presence of stable intermediate structures, it is typically impossible to assemble the target structure completely at any temperature where nucleation is rate controlling.

\begin{figure*}[t!]
  \centering
  \includegraphics[width=17.5cm]{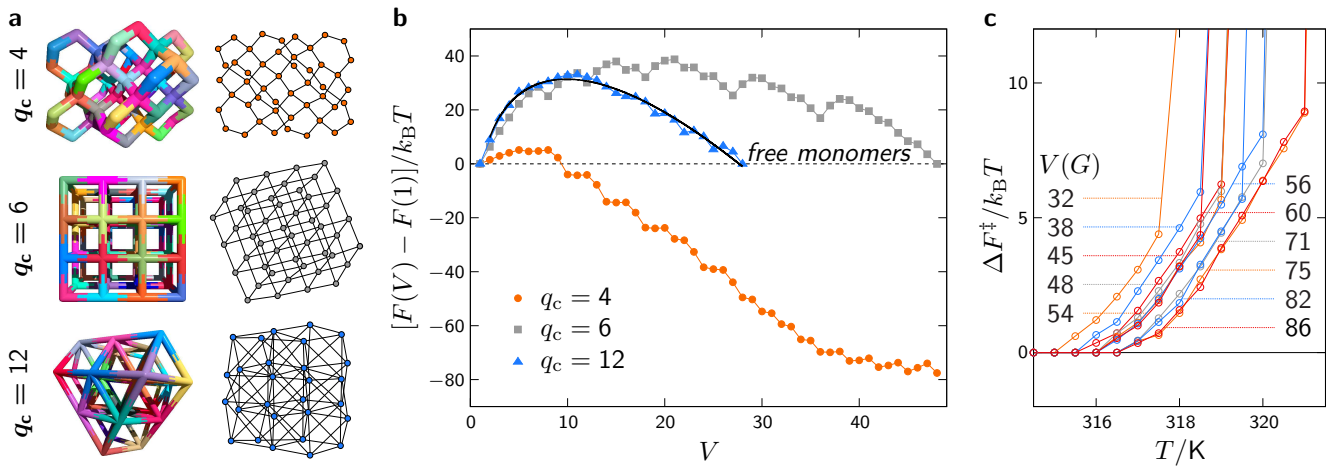}
  \caption{Dependence of the nucleation barrier on the coordination number of the target structure.  (a)~Example structures with tetrahedral coordination (${\qc = 4}$), octahedral coordination (${\qc = 6}$) and close-packed coordination (${\qc = 12}$), along with their associated connectivity graphs.  (b)~Free-energy profiles of these three structures at 50\% yield assuming identical bond energies within each structure.  The black line shows the fit of classical nucleation theory to the nucleation barrier of the ${\qc = 12}$ structure.  (c)~The dependence of the nucleation barrier on the total number of strands, $V(G)$, in DNA-brick structures with randomly chosen DNA sequences.  The nucleation temperature does not increase monotonically with $V(G)$ in these roughly cuboidal structures since surface effects are considerable.}
  \label{fig:coordination_number}
\end{figure*}

In order to examine the importance of a nucleation barrier for preventing misassembly, we calculate the free-energy difference between all off-pathway intermediates and all on-pathway intermediates, $\Delta \Fin$, by estimating the probability of incidental interactions between partially assembled structures.\cite{jacobs2015theoretical}
From the connectivity graph of the example DNA-brick structure, we can calculate the total free energy of aggregated clusters by considering all the ways that partially assembled structures can interact via the dangling ends of the single-stranded bricks, as shown in \figref{fig:ramp_example}c.
We also estimate this free-energy difference in the case of slow nucleation, $\Delta \Fin^\dagger$, by only allowing one of the interacting clusters in a misassembled intermediate to have ${V > V^\ddagger}$.

The above analysis supports our claim that a substantial nucleation barrier is essential for accurate self-assembly.
Our calculations show that even with very weak incidental interactions, incorrect bonding between the multiple dangling ends of large partial structures prevents error-free assembly at equilibrium, since ${\Delta \Fin > 0}$.
The presence of a nucleation barrier slows the approach to equilibrium, maintaining the viability of the correctly assembled clusters.

These theoretical predictions are confirmed by extensive Monte Carlo simulations of the structure shown in \figref{fig:ramp_example}a.
In these simulations, the DNA bricks are modeled as rigid particles that move on a cubic lattice, but otherwise the sequence complementarity and the hybridization free energies of the experimental system are preserved.\cite{reinhardt2014numerical}
Using realistic dynamics,\cite{whitelam2007avoiding} we simulate the assembly of the target structure using a single copy of each monomer.
In \figref{fig:ramp_example}f, we compare a representative trajectory from a simulation using a linear temperature ramp with a trajectory from a constant-temperature simulation starting from free monomers in solution.
We also report the largest stable cluster size averaged over many such trajectories in \figref{fig:ramp_example}g.
Nucleation first occurs within the predicted nucleation window where ${\Delta F^\ddagger \simeq 8~\kB T}$.
At 319~K, the size of the largest stable cluster coincides precisely with the predicted average cluster size at the free-energy minimum in \figref{fig:ramp_example}d.\footnote{Incomplete assembly at a constant temperature has also been observed in experiments\cite{sobczak2012rapid} as well as in our simulations of more complex structures, such as a rectangular slab with a raised checkerboard pattern.}
Intermediate structures assembled via a temperature ramp continue to grow at lower temperatures, while clusters formed directly from a solution of free monomers become arrested in conformations that are incompatible with further growth (\figref{fig:ramp_example}f,\textit{inset}).
In agreement with our theoretical predictions, the simulation results demonstrate that a time-dependent protocol is essential for correctly assembling a complete DNA-brick structure.

\section*{Coordination number controls the nucleation barrier}

In the modular assemblies reported in \refcite{ke2012three}, the maximum coordination number of bricks in the interior of the structure is four.
However, one can envisage other building blocks, such as functionalized molecular constructs or nano-colloids, that have a different coordination number.
To investigate the effect of the coordination number on the nucleation barrier, we compare the free-energy profile of a 48-strand DNA-brick structure with those of two higher-coordinated structures (\figref{fig:coordination_number}a): a simple cubic structure with coordination number ${\qc = 6}$ and a close-packed structure with ${\qc = 12}$.  (For a discussion of two-dimensional structures, see \secref{sec:2d}.)
In \figref{fig:coordination_number}b, we show the free-energy profiles at 50\% yield assuming identical bond energies within each structure.

\begin{figure*}[t!]
  \centering
  \includegraphics[width=17.5cm]{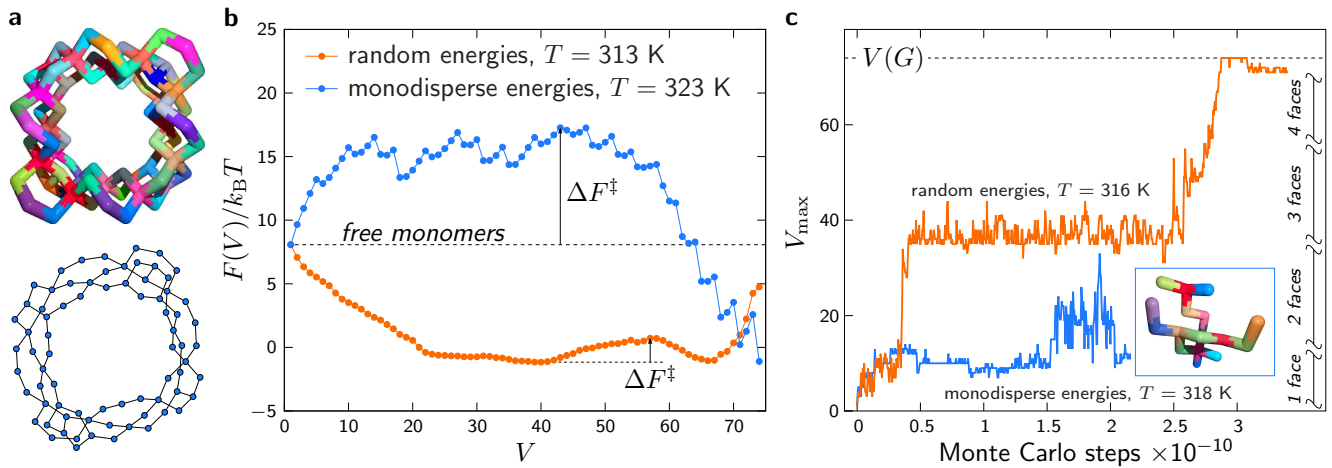}
  \caption{Random DNA sequences can improve the kinetics of self-assembly.  (a)~A non-convex, 74-strand DNA-brick structure and its associated connectivity graph.  (b)~Free-energy profiles and effective nucleation barriers, $\Delta F^\ddagger$, with both random and monodisperse designed interaction strengths.   The temperatures are chosen such that the mean hybridization free energies of these two distributions are equal.  (c)~Representative lattice Monte Carlo simulation trajectories with heterogeneous and monodisperse bond energies.  Correct assembly proceeds by assembling one face of the structure at a time, as shown on the right.  The temperature of the monodisperse system must be lowered to achieve nucleation, but the largest cluster quickly becomes kinetically trapped (\textit{inset}).}
  \label{fig:central_hole}
\end{figure*}

One striking difference between the ${\qc = 4}$ structure and the higher-coordinated examples is the stability of the target at 50\% yield.
In the DNA-brick structure, the target structure coexists in nearly equal populations with a partial structure that is missing a single cycle.
In the structures with higher coordination numbers, however, the target has the same free energy as the free monomers at 50\% yield.
Intermediate structures are therefore globally unstable at all temperatures, as predicted by CNT.

A second point of distinction among these structures lies in the relative stability of intermediate cluster sizes.
Whereas the DNA-brick structure assembles by completing individual cycles, the cubic structure grows by adding one face at a time to an expanding cuboid.\footnote{This equilibrium growth pathway is similar to that described in \refcite{shneidman2003lowest}.}
With ${\qc = 12}$, the greater diversity of fragments with the same number of vertices smooths out the free-energy profile near the top of the nucleation barrier.
The fitted black line in \figref{fig:coordination_number}b shows that the assembly of this structure does in fact obey CNT (see \secref{sec:cnt}).

The differences among these free-energy profiles originate from the topologies of the connectivity graphs of the example structures.
The most important determinant of the nucleation behavior is simply the number of vertices required to complete each additional cycle in the target connectivity graph, which is controlled by the maximum coordination number of the subunits.\footnote{Although the coordination number also affects the rotational entropy in lattice-based simulations, altering the rotational entropy in our theoretical method has the same effect as changing the monomer concentration (see \secref{sec:concentration}) and thus does not affect the shape of the free-energy profile.}
Our findings imply that controlled self-assembly of three-dimensional addressable structures is unlikely to be achieved straightforwardly using subunits with coordination numbers higher than four.
In higher-coordinated structures, which are well described by CNT, it would be necessary to go to high supersaturation in order to find a surmountable nucleation barrier; however, such an approach is likely to fail due to kinetic trapping.\cite{whitelam2014statistical}
Yet in DNA-brick structures, the nucleation barrier is surmountable at low supersaturation and is relatively insensitive to the size of the target structure (\figref{fig:coordination_number}c).
The reliable self-assembly of large DNA-brick structures is thus a direct consequence of the small number of bonds made by each brick.

\section*{Heterogeneous bond energies improve kinetics}

Recent publications have argued that equal bond energies should enhance the stability of the designed structure\cite{hormoz2011design} and reduce errors during growth.\cite{hedges2014growth}
By contrast, we find that the kinetics of DNA-brick assembly are actually worse if one selects DNA sequences that minimize the variance in the bond energies.
Our observation is consistent with the successful use of random DNA sequences in the original experiments with DNA bricks.\cite{wei2012complex,ke2012three}
Here again, the nucleation behavior is responsible for this unexpected result.

To demonstrate the difference between random DNA sequences and sequences chosen to yield monodisperse bond energies, we consider the relatively simple non-convex DNA-brick structure shown in \figref{fig:central_hole}a.
This 74-brick structure, constructed by removing the interior strands and two faces from a cuboidal structure, assembles roughly face-by-face when using random DNA sequences.
The relevant nucleation barrier, as predicted theoretically in \figref{fig:central_hole}b and confirmed with Monte Carlo simulations in \figref{fig:central_hole}c, is the completion of the third face.
With monodisperse bond energies and an equivalent mean interaction strength, a much larger nucleation barrier appears before the first face forms.
Attempts to reduce this nucleation barrier by increasing the mean bond energy result in kinetic trapping and arrested growth.
Despite promising fluctuations in the largest cluster size in the simulation trajectory with monodisperse energies, multiple competing nuclei appear, and the largest cluster remains poorly configured for further assembly (\figref{fig:central_hole}c,\textit{inset}).

The use of sequences with a broad distribution of hybridization free energies results in a more suitable nucleation barrier because such a distribution selectively stabilizes small and floppy intermediate structures.
This is a statistical effect: since there are far fewer ways of constructing a maximally connected fragment with a given number of monomers, the chance that randomly assigned sequences concentrate the strongest bonds in a compact fragment is vanishingly small in a large structure.
As a result, the dominant nucleation pathways no longer need to follow the maximally connected fragments.
The use of a broad distribution of bond energies therefore tends to reduce nucleation barriers, since unstable fragments near the top of a barrier contain fewer cycles and are thus affected more significantly by the variance in the bond-energy distribution.

\section*{Design principles for (DNA) nanostructures}

The insights provided by our predictive theory allow us to understand the general principles underlying the unexpected success of DNA-brick self-assembly.
Slow, controlled nucleation at low supersaturation is achieved for large structures since each brick can only make a small number of designed connections.
Because of an appreciable nucleation barrier that appears in a narrow temperature window, monomer depletion does not pose a significant problem for one-pot assembly.
Surprisingly, complex structures with randomly selected complementary DNA sequences experience enhanced nucleation, making larger intermediate structures kinetically accessible at higher temperatures.

The use of a temperature ramp plays a more crucial role than previously thought.
Cooling the DNA-brick solution slowly is not just a convenient way of locating good assembly conditions, as in the case of conventional crystals; rather, it is an essential non-equilibrium protocol for achieving error-free assembly of finite-sized structures.
The explanation of slow nucleation and fast growth that was originally proposed in \refscite{wei2012complex} and \onlinecite{ke2012three} is therefore incomplete: fast growth allows the DNA bricks to assemble into a stable, on-pathway intermediate that must be annealed at lower temperatures to complete the target structure.
Remaining out of equilibrium throughout the assembly protocol, as is necessary in order to avoid the aggregation of partial structures, relies on the slow diffusion of large intermediates.
This is a reasonable assumption, since the rate of diffusion changes approximately inversely with the radius of a fragment in solution.\cite{einstein1905molekularkinetischen}

Our approach also suggests how to improve the design of DNA-brick nanostructures beyond the random selection of uniformly distributed DNA sequences.
For a given target structure, it is easy to tune the nucleation barrier by adjusting the statistical distribution of bond energies.
Complementary DNA sequences can then be assigned to the structure in order to achieve the desired distribution of hybridization free energies.
Furthermore, with an understanding of the origin of the nucleation barrier in a particular structure, it is possible to optimize the annealing protocol rationally in order to increase the yield of the target assembly.
Our approach also provides a means of systematically investigating how local modifications to the coordination number through the fusing of adjacent strands affect the nucleation behavior of DNA-brick structures.\cite{li2014controlled}

The theoretical method used here greatly simplifies the quantitative prediction of nucleation barriers and intermediate structures with widespread applications for controlling the self-assembly of biomolecular or synthetic building blocks.
Addressable self-assembly holds great promise for building intricate three-dimensional structures that are likely to require optimization on a case-by-case basis.
Because our predictive theory is sensitive to the details of a particular target structure, performing these calculations for nanostructures of experimental interest will enable the precise engineering of assembly properties at the design stage.
In order for potential users to perform such experimental protocol design, we provide a user-friendly software package online at \url{https://github.com/wmjac/pygtsa}.

\begin{acknowledgments}
This work was carried out with support from the European Research Council (Advanced Grant 227758) and the Engineering and Physical Sciences Research Council Programme Grant EP/I001352/1.
W.M.J.~acknowledges support from the Gates Cambridge Trust and the National Science Foundation Graduate Research Fellowship under Grant No.~DGE-1143678.
\end{acknowledgments}

\section*{Methods}

\subsection*{DNA hybridization free energies}
{\small
We compute the hybridization free energies of complementary 8-nucleotide DNA sequences using established empirical formulae\cite{santalucia2004thermodynamics,koehler2005thermodynamic} assuming salt concentrations of [Na$^+$]~=~1~mol\;dm$^{-3}$ and [Mg$^{2+}$]~=~0.08~mol\;dm$^{-3}$.
For the calculations with monodisperse bond energies, we use the sequences provided in \refcite{hedges2014growth}.
The strengths of incidental interactions are estimated based on the longest attractive overlap for each pair of non-complementary sequences.
In calculations of the equilibrium yield and free-energy profiles, we report the average thermodynamic properties using 1000 randomly chosen complete sets of DNA sequences.
See \secref{sec:distributions} for further details.
}

\subsection*{Lattice Monte Carlo simulations}
{\small
Constant-temperature lattice Monte Carlo simulations are carried out using the virtual move Monte Carlo algorithm\cite{whitelam2007avoiding} in order to produce physical dynamics.
Rigid particles, each with four distinct patches fixed in a tetrahedral arrangement, are confined to a cubic lattice.
A single copy of each required subunit is present in the simulation box with $62\times62\times62$ lattice sites.
Complete details are given in \refcite{reinhardt2014numerical}.
For comparison with the results of these simulations, the theoretical calculations reported here assume the same dimensionless monomer concentration, ${\rho = 62^{-3}}$, lattice coordination number, ${\qc = 4}$, and fixed number of dihedral angles, ${\qd = 3}$ (see \secref{sec:theory}).
}

%


\renewcommand{\thefigure}{S\arabic{figure}}
\renewcommand{\thesection}{S\arabic{section}}
\setcounter{figure}{0}
\setcounter{section}{0}

\clearpage

\onecolumngrid
\section*{Supplementary Information}
\twocolumngrid

\section{Theoretical free-energy and equilibrium yield calculations}
\label{sec:theory}

We construct the connectivity graph, $G$, from the designed bonds between adjacent subunits in the target structure.
An example connectivity graph is shown in \figref{fig:ramp_example}b.
From this graph we are able to determine all the relevant thermodynamic properties of the intermediate and target structures in a near-equilibrium assembly protocol.
A thorough explanation of this theoretical method is presented in \refcite{jacobs2015theoretical}; here we summarize the key equations.

The connected subgraphs (`fragments') of the target connectivity graph are grouped into sets, $h(E,V)$, in which all fragments have precisely $E$ edges and $V$ vertices.
Assuming that subunits can only form clusters with designed bonds, the dimensionless grand potential, $-\ln\Xi$, can be written in terms of a sum over all sets of fragments,
\begin{equation}
  \label{eq:lnXi}
  \Zid \equiv \ln \Xi = \sum_{E,V} | h(E,V) | \,\barzEV.
\end{equation}
The average fugacity of the fragments in each set, $\barzEV$, depends on the topologies of the fragment graphs as well as the geometry of the subunits and the solution conditions.
Ignoring excluded volume interactions, we can approximate $\ln\barzEV$ as
\begin{eqnarray}
  \label{eq:lnzEV}
  \ln\barzEV &=& E \bteEV + V \ln \rho \nonumber \\
  &\quad& - (V - 1) \ln \qc - (V - \barBEV - 1) \ln \qd,
\end{eqnarray}
where $\kB\ln\qc$ is the rotational entropy of a monomer, $\kB\ln\qd$ is the dihedral entropy of an unconstrained dimer and $\rho$ is the dimensionless concentration.
The mean dimensionless dihedral entropy of a fragment is $\barBEV \ln\qd \equiv \ln \left\langle \qd^{B(g)} \right\rangle_{g \in h(E,V)}$, where $B(g)$ is the number of bridges\cite{bondy1976graph} in the fragment $g$.
The exponentially weighted mean bond energy within each set, $\bteEV$, is
\begin{equation}
  \label{eq:mean_energies}
  \bteEV \equiv \left.\mathlarger{\Biggl\langle} \frac{1}{E} \ln \left\langle \exp \! \sum_{b \in \mathcal{E}(g)} \!\! \be_b \right\rangle_{\!\!\! g \in h(E,V)} \mathlarger{\Biggr\rangle}\right._{\!\!\!\! \av} \!\!\!\!,
\end{equation}
where $\epsilon_b$ is the hybridization free energy of bond $b$, ${\beta \equiv 1/\kB T}$ is the inverse temperature and $\mathcal{E}(g)$ is the edge set of fragment $g$.
The inner average runs over all fragments in the set $h(E,V)$ with \textit{quenched} random bond energies $\{\epsilon_b\}$.
In the case of DNA-brick structures, the outer average samples DNA sequences so that each complete set of bond energies for the target structure is chosen independently from the same distribution of hybridization free energies.

The free energy of a correctly bonded cluster of $V$ monomers is thus
\begin{equation}
  \beta F(V) \equiv -\ln \sum_E | h(E,V) | \,\barzEV,
\end{equation}
since we do not distinguish among equally sized clusters with varying compositions.
This definition is appropriate for studying nucleation, as any subset of monomers has the potential to serve as a nucleation site.

The equilibrium yield, $\eta$, is defined as the fraction of all clusters in solution that are correctly formed,
\begin{equation}
  \label{eq:equilibrium_yield}
  \eta \equiv \frac{\langle N_G \rangle}{\sum_g \langle N_g \rangle} = \frac{z_G}{\Zid},
\end{equation}
where $\langle N_G \rangle$ is the grand-canonical average number of copies of fragment $g$ in solution and $z_G$ is the fugacity of the target structure.
In the case of structures with higher coordination numbers, a few edges may be removed from the connectivity graph without allowing any subunit to disassociate or rotate.
For these structures, we replace $z_G$ in \eqref{eq:equilibrium_yield} with a sum over the fugacities of all fragments that enforce the correct geometry of the target structure.

\section{Classical nucleation theory}
\label{sec:cnt}

Classical nucleation theory predicts a free-energy barrier for the nucleation of a stable, ordered structure from an unstable, fluid phase.
The height of the barrier and the size of the critical nucleus vary with the degree of supersaturation of the fluid phase.\cite{oxtoby1992homogeneous,sear2007nucleation}
Assuming spherical nuclei, the classical prediction for the free-energy difference between a nucleus of the ordered phase containing $V$ monomers and the bulk fluid phase is
\begin{equation}
  \Delta F(V) = -\Delta \mu V + \gamma \bigl( 36 \pi \rho_{\text{ord}}^{-2} \bigr)^{1/3} \bigl( V - V_0 \bigr)^{2/3},~
  \label{eq:cnt}
\end{equation}
where $\Delta \mu$ is the bulk free-energy difference per particle between the fluid phase and the ordered phase, $\gamma$ is the free-energy cost per unit area of forming an interface between the two phases and $\rho_{\text{ord}}$ is the number density of the ordered phase.
In \figref{fig:coordination_number}b, the black line shows the fit to \eqref{eq:cnt} with ${V_0 = 1}$.

\section{Concentration dependence}
\label{sec:concentration}

\begin{figure}[t]
  \centering
  \includegraphics[width=8.5cm]{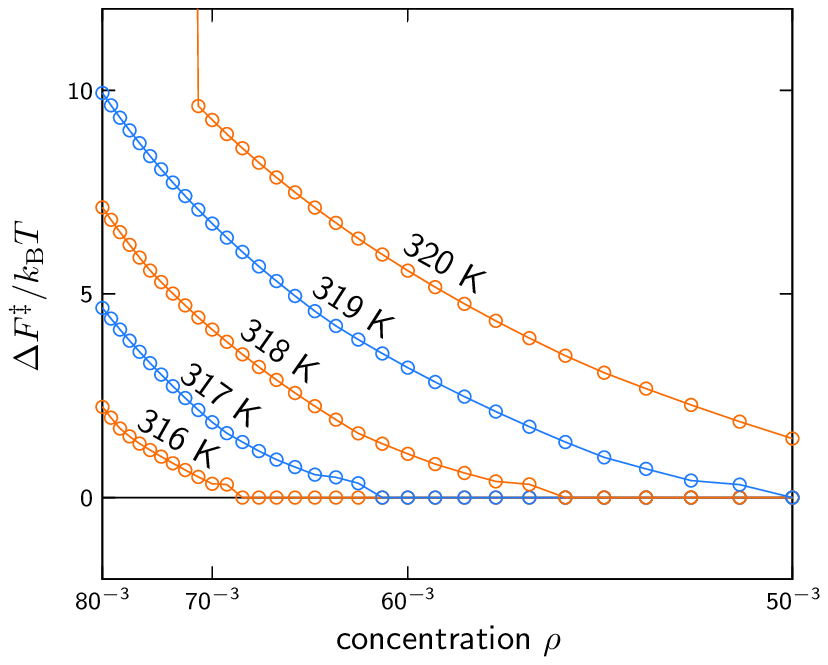}
  \caption{The dependence of the height of the nucleation barrier, $\Delta F^\ddagger$, on the dimensionless concentration, $\rho$, for the example 86-strand structure shown in \figref{fig:ramp_example}a.}
  \label{fig:concentration_dependence}
\end{figure}

In the main text, we report all results of the theoretical calculations assuming a dimensionless concentration of $62^{-3}$ for comparison with the lattice Monte Carlo simulations.
Changing this concentration shifts both the equilibrium yield and the nucleation barrier linearly with $\ln\rho$.
The concentration dependence of the nucleation barrier of the example 86-strand structure at several temperatures is shown in \figref{fig:concentration_dependence}.
Although we have assumed equal concentrations of all monomers, polydispersity in the monomer concentrations can be easily incorporated into the theoretical treatment in much the same way as the distributions of designed interaction energies.

\begin{figure*}[t]
  \centering
  \includegraphics[width=12cm]{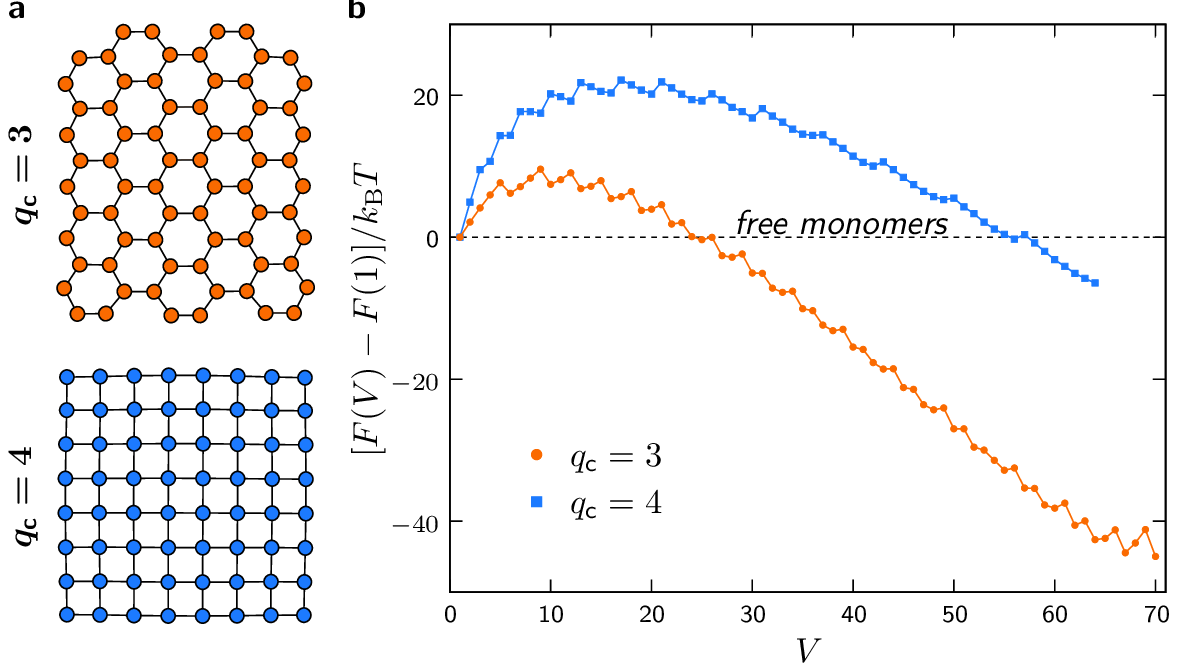}
  \caption{Dependence of the nucleation barrier on the coordination number of two-dimensional structures.  (a) Connectivity graphs of two-dimensional structures with coordination numbers ${\qc = 3}$ and ${\qc = 4}$.  (b) The free-energy profiles of these structures at 50\% yield assuming identical bond energies within each structure.  For both structures, we assume ${\qd = 2}$.}
  \label{fig:two_dimensional_structures}
\end{figure*}

\section{Assembly of two-dimensional structures}
\label{sec:2d}

We find that the shapes of the free-energy profiles of two-dimensional structures are also strongly affected by their coordination numbers.
In \figref{fig:two_dimensional_structures}, we show the free-energy profiles at 50\% yield of two similarly sized two-dimensional structures with coordination numbers ${\qc = 3}$ and ${\qc = 4}$.
The behavior of the two-dimensional structure with ${\qc = 3}$ is similar to that of the four-coordinated three-dimensional structures examined in the main text.
The two-dimensional structure with ${\qc = 4}$ exhibits the same face-by-face assembly as three-dimensional structures with octahedral coordination; in the two-dimensional case, however, coexistence at 50\% yield occurs between the target structure and fragments with the same number of monomers but fewer bonds.

Two-dimensional DNA-tile structures with ${\qc = 4}$ have been successfully assembled.\cite{wei2012complex}
In general, the nucleation barriers in two-dimensional structures are much lower than in three-dimensional structures with a similar number of monomers.
Consequently, a lower supersaturation is required in order for the target structure to become kinetically accessible, and kinetic trapping is therefore less likely to interfere with accurate assembly.
Nevertheless, these calculations suggest that lower coordinated two-dimensional structures, such as the hexagonal lattice pictured in \figref{fig:two_dimensional_structures}a, might assemble more robustly in experiments.

\section{Hybridization free-energy distributions}
\label{sec:distributions}

\begin{figure*}[t]
  \centering
  \includegraphics[width=12cm]{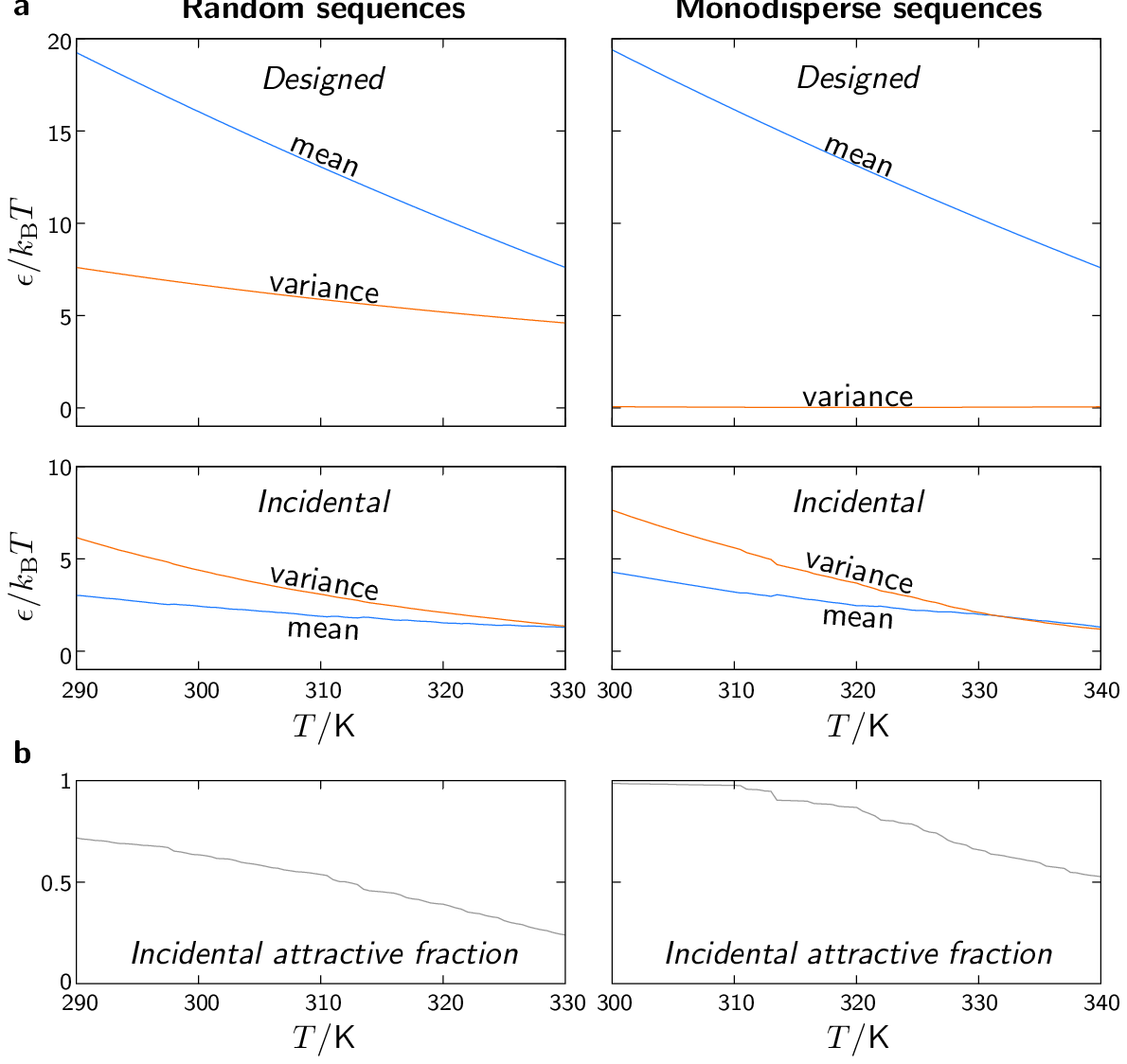}
  \caption{DNA hybridization free-energy distributions.  (a) The mean and variance of the hybridization free-energy distributions with randomly chosen DNA sequences (left) and with sequences selected to yield monodisperse bond energies (right).  Designed interactions refer to complementary DNA sequences, while incidental interactions report the hybridization free energy of the most attractive overlapping region of two non-complementary sequences.  Note that the means of the designed interactions of these two distributions are shifted by approximately 10~K.  (b) The fraction of pairs of non-complementary sequences that attract more strongly than 100~K.}
  \label{fig:hybridization_free_energies}
\end{figure*}

DNA hybridization free energies are strongly temperature-dependent.\cite{santalucia2004thermodynamics,koehler2005thermodynamic}
In \figref{fig:hybridization_free_energies}a, we compare the two hybridization free-energy distributions used in the main text.
The mean and the variance of the hybridization free energies of 8-nucleotide sequences are shown for both the case of randomly chosen sequences and the case of sequences selected to yield monodisperse bond energies.\cite{hedges2014growth}
Designed interactions occur between complementary sequences, while incidental interactions are calculated based on the most attractive overlapping regions of two non-complementary sequences.

In the lattice Monte Carlo simulations, all monomers on adjacent lattice sites experience a weak 100~K repulsion at all temperatures.
In calculations involving designed interactions, this repulsion is subtracted from the mean interaction strength.
When estimating incidental interactions, we ignore associations between non-complementary sequences that have a maximum attractive interaction of less than 100~K.
The means and variances reported in \figref{fig:hybridization_free_energies}a are thus calculated based on the fraction of pairs of non-complementary sequences that attract more strongly than 100~K; this fraction is shown in \figref{fig:hybridization_free_energies}b.

These incidental interaction distributions are clearly approximate and are defined in order to match the lattice Monte Carlo simulations.
Nevertheless, the choices made in defining these distributions have a negligible effect on the calculated values of $\Zin$\cite{jacobs2015theoretical} and are irrelevant to the prediction of nucleation barriers and equilibrium yields.
In order to apply this theoretical method to an experimental system, the designed interaction distributions should be recalculated in accordance with the experimental solution conditions.

\begin{thebibliography}{29}%
\makeatletter
\providecommand \@ifxundefined [1]{%
 \@ifx{#1\undefined}
}%
\providecommand \@ifnum [1]{%
 \ifnum #1\expandafter \@firstoftwo
 \else \expandafter \@secondoftwo
 \fi
}%
\providecommand \@ifx [1]{%
 \ifx #1\expandafter \@firstoftwo
 \else \expandafter \@secondoftwo
 \fi
}%
\providecommand \natexlab [1]{#1}%
\providecommand \enquote  [1]{``#1''}%
\providecommand \bibnamefont  [1]{#1}%
\providecommand \bibfnamefont [1]{#1}%
\providecommand \citenamefont [1]{#1}%
\providecommand \href@noop [0]{\@secondoftwo}%
\providecommand \href [0]{\begingroup \@sanitize@url \@href}%
\providecommand \@href[1]{\@@startlink{#1}\@@href}%
\providecommand \@@href[1]{\endgroup#1\@@endlink}%
\providecommand \@sanitize@url [0]{\catcode `\\12\catcode `\$12\catcode
  `\&12\catcode `\#12\catcode `\^12\catcode `\_12\catcode `\%12\relax}%
\providecommand \@@startlink[1]{}%
\providecommand \@@endlink[0]{}%
\providecommand \url  [0]{\begingroup\@sanitize@url \@url }%
\providecommand \@url [1]{\endgroup\@href {#1}{\urlprefix }}%
\providecommand \urlprefix  [0]{URL }%
\providecommand \Eprint [0]{\href }%
\providecommand \doibase [0]{http://dx.doi.org/}%
\providecommand \selectlanguage [0]{\@gobble}%
\providecommand \bibinfo  [0]{\@secondoftwo}%
\providecommand \bibfield  [0]{\@secondoftwo}%
\providecommand \translation [1]{[#1]}%
\providecommand \BibitemOpen [0]{}%
\providecommand \bibitemStop [0]{}%
\providecommand \bibitemNoStop [0]{.\EOS\space}%
\providecommand \EOS [0]{\spacefactor3000\relax}%
\providecommand \BibitemShut  [1]{\csname bibitem#1\endcsname}%
\let\auto@bib@innerbib\@empty
\bibitem [{\citenamefont {Wei}, \citenamefont {Dai},\ and\ \citenamefont
  {Yin}(2012)}]{wei2012complex}%
  \BibitemOpen
  \bibfield  {author} {\bibinfo {author} {\bibfnamefont {B.}~\bibnamefont
  {Wei}}, \bibinfo {author} {\bibfnamefont {M.}~\bibnamefont {Dai}}, \ and\
  \bibinfo {author} {\bibfnamefont {P.}~\bibnamefont {Yin}},\ }\href@noop {}
  {\bibfield  {journal} {\bibinfo  {journal} {Nature}\ }\textbf {\bibinfo
  {volume} {485}},\ \bibinfo {pages} {623} (\bibinfo {year}
  {2012})}\BibitemShut {NoStop}%
\bibitem [{\citenamefont {Ke}\ \emph {et~al.}(2012)\citenamefont {Ke},
  \citenamefont {Ong}, \citenamefont {Shih},\ and\ \citenamefont
  {Yin}}]{ke2012three}%
  \BibitemOpen
  \bibfield  {author} {\bibinfo {author} {\bibfnamefont {Y.}~\bibnamefont
  {Ke}}, \bibinfo {author} {\bibfnamefont {L.~L.}\ \bibnamefont {Ong}},
  \bibinfo {author} {\bibfnamefont {W.~M.}\ \bibnamefont {Shih}}, \ and\
  \bibinfo {author} {\bibfnamefont {P.}~\bibnamefont {Yin}},\ }\href@noop {}
  {\bibfield  {journal} {\bibinfo  {journal} {Science}\ }\textbf {\bibinfo
  {volume} {338}},\ \bibinfo {pages} {1177} (\bibinfo {year}
  {2012})}\BibitemShut {NoStop}%
\bibitem [{\citenamefont {Rothemund}\ and\ \citenamefont
  {Winfree}(2000)}]{rothemund2000program}%
  \BibitemOpen
  \bibfield  {author} {\bibinfo {author} {\bibfnamefont {P.~W.}\ \bibnamefont
  {Rothemund}}\ and\ \bibinfo {author} {\bibfnamefont {E.}~\bibnamefont
  {Winfree}},\ }in\ \href@noop {} {\emph {\bibinfo {booktitle} {Proceedings of
  the Thirty-second Annual {ACM} Symposium on Theory of Computing}}}\ (\bibinfo
  {organization} {ACM},\ \bibinfo {year} {2000})\ pp.\ \bibinfo {pages}
  {459--468}\BibitemShut {NoStop}%
\bibitem [{\citenamefont {Zhang}\ \emph {et~al.}(2013)\citenamefont {Zhang},
  \citenamefont {Song}, \citenamefont {Besenbacher}, \citenamefont {Dong},\
  and\ \citenamefont {Gothelf}}]{zhang2013self}%
  \BibitemOpen
  \bibfield  {author} {\bibinfo {author} {\bibfnamefont {Z.}~\bibnamefont
  {Zhang}}, \bibinfo {author} {\bibfnamefont {J.}~\bibnamefont {Song}},
  \bibinfo {author} {\bibfnamefont {F.}~\bibnamefont {Besenbacher}}, \bibinfo
  {author} {\bibfnamefont {M.}~\bibnamefont {Dong}}, \ and\ \bibinfo {author}
  {\bibfnamefont {K.~V.}\ \bibnamefont {Gothelf}},\ }\href@noop {} {\bibfield
  {journal} {\bibinfo  {journal} {Angew. Chem. Int. Ed. Engl.}\ }\textbf
  {\bibinfo {volume} {52}},\ \bibinfo {pages} {9219} (\bibinfo {year}
  {2013})}\BibitemShut {NoStop}%
\bibitem [{\citenamefont {Wei}\ \emph {et~al.}(2013)\citenamefont {Wei},
  \citenamefont {Dai}, \citenamefont {Myhrvold}, \citenamefont {Ke},
  \citenamefont {Jungmann},\ and\ \citenamefont {Yin}}]{wei2013design}%
  \BibitemOpen
  \bibfield  {author} {\bibinfo {author} {\bibfnamefont {B.}~\bibnamefont
  {Wei}}, \bibinfo {author} {\bibfnamefont {M.}~\bibnamefont {Dai}}, \bibinfo
  {author} {\bibfnamefont {C.}~\bibnamefont {Myhrvold}}, \bibinfo {author}
  {\bibfnamefont {Y.}~\bibnamefont {Ke}}, \bibinfo {author} {\bibfnamefont
  {R.}~\bibnamefont {Jungmann}}, \ and\ \bibinfo {author} {\bibfnamefont
  {P.}~\bibnamefont {Yin}},\ }\href@noop {} {\bibfield  {journal} {\bibinfo
  {journal} {J. Am. Chem. Soc.}\ }\textbf {\bibinfo {volume} {135}},\ \bibinfo
  {pages} {18080} (\bibinfo {year} {2013})}\BibitemShut {NoStop}%
\bibitem [{\citenamefont {Rothemund}\ and\ \citenamefont
  {Andersen}(2012)}]{rothemund2012nanotechnology}%
  \BibitemOpen
  \bibfield  {author} {\bibinfo {author} {\bibfnamefont {P.~W.}\ \bibnamefont
  {Rothemund}}\ and\ \bibinfo {author} {\bibfnamefont {E.~S.}\ \bibnamefont
  {Andersen}},\ }\href@noop {} {\bibfield  {journal} {\bibinfo  {journal}
  {Nature}\ }\textbf {\bibinfo {volume} {485}},\ \bibinfo {pages} {584}
  (\bibinfo {year} {2012})}\BibitemShut {NoStop}%
\bibitem [{\citenamefont {Gothelf}(2012)}]{gothelf2012lego}%
  \BibitemOpen
  \bibfield  {author} {\bibinfo {author} {\bibfnamefont {K.~V.}\ \bibnamefont
  {Gothelf}},\ }\href@noop {} {\bibfield  {journal} {\bibinfo  {journal}
  {Science}\ }\textbf {\bibinfo {volume} {338}},\ \bibinfo {pages} {1159}
  (\bibinfo {year} {2012})}\BibitemShut {NoStop}%
\bibitem [{\citenamefont {Kim}\ \emph {et~al.}(2008)\citenamefont {Kim},
  \citenamefont {Scarlett}, \citenamefont {Biancaniello}, \citenamefont
  {Sinno},\ and\ \citenamefont {Crocker}}]{kim2008probing}%
  \BibitemOpen
  \bibfield  {author} {\bibinfo {author} {\bibfnamefont {A.~J.}\ \bibnamefont
  {Kim}}, \bibinfo {author} {\bibfnamefont {R.}~\bibnamefont {Scarlett}},
  \bibinfo {author} {\bibfnamefont {P.~L.}\ \bibnamefont {Biancaniello}},
  \bibinfo {author} {\bibfnamefont {T.}~\bibnamefont {Sinno}}, \ and\ \bibinfo
  {author} {\bibfnamefont {J.~C.}\ \bibnamefont {Crocker}},\ }\href@noop {}
  {\bibfield  {journal} {\bibinfo  {journal} {Nat. Mater.}\ }\textbf {\bibinfo
  {volume} {8}},\ \bibinfo {pages} {52} (\bibinfo {year} {2008})}\BibitemShut
  {NoStop}%
\bibitem [{\citenamefont {Reinhardt}\ and\ \citenamefont
  {Frenkel}(2014)}]{reinhardt2014numerical}%
  \BibitemOpen
  \bibfield  {author} {\bibinfo {author} {\bibfnamefont {A.}~\bibnamefont
  {Reinhardt}}\ and\ \bibinfo {author} {\bibfnamefont {D.}~\bibnamefont
  {Frenkel}},\ }\href@noop {} {\bibfield  {journal} {\bibinfo  {journal} {Phys.
  Rev. Lett.}\ }\textbf {\bibinfo {volume} {112}},\ \bibinfo {pages} {238103}
  (\bibinfo {year} {2014})}\BibitemShut {NoStop}%
\bibitem [{\citenamefont {Jacobs}, \citenamefont {Reinhardt},\ and\
  \citenamefont {Frenkel}(2015)}]{jacobs2015theoretical}%
  \BibitemOpen
  \bibfield  {author} {\bibinfo {author} {\bibfnamefont {W.~M.}\ \bibnamefont
  {Jacobs}}, \bibinfo {author} {\bibfnamefont {A.}~\bibnamefont {Reinhardt}}, \
  and\ \bibinfo {author} {\bibfnamefont {D.}~\bibnamefont {Frenkel}},\
  }\href@noop {} {\bibfield  {journal} {\bibinfo  {journal} {J. Chem. Phys.}\
  }\textbf {\bibinfo {volume} {142}},\ \bibinfo {pages} {021101} (\bibinfo
  {year} {2015})}\BibitemShut {NoStop}%
\bibitem [{Note1()}]{Note1}%
  \BibitemOpen
  \bibinfo {note} {We note that thermodynamically stable partial structures
  have also been observed in previously reported simulations\cite
  {reinhardt2014numerical} of DNA-brick structures consisting of approximately
  1,000 distinct subunits.}\BibitemShut {Stop}%
\bibitem [{\citenamefont {Gibbs}(1878)}]{gibbs1878equilibrium}%
  \BibitemOpen
  \bibfield  {author} {\bibinfo {author} {\bibfnamefont {J.~W.}\ \bibnamefont
  {Gibbs}},\ }\href@noop {} {\bibfield  {journal} {\bibinfo  {journal} {Am. J.
  Sci.}\ }\textbf {\bibinfo {volume} {96}},\ \bibinfo {pages} {441} (\bibinfo
  {year} {1878})}\BibitemShut {NoStop}%
\bibitem [{\citenamefont {Oxtoby}(1992)}]{oxtoby1992homogeneous}%
  \BibitemOpen
  \bibfield  {author} {\bibinfo {author} {\bibfnamefont {D.~W.}\ \bibnamefont
  {Oxtoby}},\ }\href@noop {} {\bibfield  {journal} {\bibinfo  {journal} {J.
  Phys. Cond. Matter}\ }\textbf {\bibinfo {volume} {4}},\ \bibinfo {pages}
  {7627} (\bibinfo {year} {1992})}\BibitemShut {NoStop}%
\bibitem [{\citenamefont {Sear}(2007)}]{sear2007nucleation}%
  \BibitemOpen
  \bibfield  {author} {\bibinfo {author} {\bibfnamefont {R.~P.}\ \bibnamefont
  {Sear}},\ }\href@noop {} {\bibfield  {journal} {\bibinfo  {journal} {J. Phys.
  Cond. Matter}\ }\textbf {\bibinfo {volume} {19}},\ \bibinfo {pages} {033101}
  (\bibinfo {year} {2007})}\BibitemShut {NoStop}%
\bibitem [{Note2()}]{Note2}%
  \BibitemOpen
  \bibinfo {note} {DNA-brick structures have also been assembled at constant
  temperature by changing the solution conditions during the course of the
  experiment.\cite {zhang2013self} For simplicity, we assume that the
  experimental control parameter is the temperature, but clearly other methods
  of altering the DNA hybridization free energies during the assembly process
  may also be suitable.}\BibitemShut {Stop}%
\bibitem [{\citenamefont {Whitelam}\ and\ \citenamefont
  {Geissler}(2007)}]{whitelam2007avoiding}%
  \BibitemOpen
  \bibfield  {author} {\bibinfo {author} {\bibfnamefont {S.}~\bibnamefont
  {Whitelam}}\ and\ \bibinfo {author} {\bibfnamefont {P.~L.}\ \bibnamefont
  {Geissler}},\ }\href@noop {} {\bibfield  {journal} {\bibinfo  {journal} {J.
  Chem. Phys.}\ }\textbf {\bibinfo {volume} {127}},\ \bibinfo {pages} {154101}
  (\bibinfo {year} {2007})}\BibitemShut {NoStop}%
\bibitem [{Note3()}]{Note3}%
  \BibitemOpen
  \bibinfo {note} {Incomplete assembly at a constant temperature has also been
  observed in experiments\cite {sobczak2012rapid} as well as in our simulations
  of more complex structures, such as a rectangular slab with a raised
  checkerboard pattern.}\BibitemShut {Stop}%
\bibitem [{Note4()}]{Note4}%
  \BibitemOpen
  \bibinfo {note} {This equilibrium growth pathway is similar to that described
  in Ref.~\protect \rev@citealpnum {shneidman2003lowest}.}\BibitemShut {Stop}%
\bibitem [{Note5()}]{Note5}%
  \BibitemOpen
  \bibinfo {note} {Although the coordination number also affects the rotational
  entropy in lattice-based simulations, altering the rotational entropy in our
  theoretical method has the same effect as changing the monomer concentration
  (see Sec.~\ref {sec:concentration}) and thus does not affect the shape of the
  free-energy profile.}\BibitemShut {Stop}%
\bibitem [{\citenamefont {Whitelam}\ and\ \citenamefont
  {Jack}(2014)}]{whitelam2014statistical}%
  \BibitemOpen
  \bibfield  {author} {\bibinfo {author} {\bibfnamefont {S.}~\bibnamefont
  {Whitelam}}\ and\ \bibinfo {author} {\bibfnamefont {R.~L.}\ \bibnamefont
  {Jack}},\ }\href@noop {} {\bibfield  {journal} {\bibinfo  {journal} {arXiv
  preprint arXiv:1407.2505}\ } (\bibinfo {year} {2014})}\BibitemShut {NoStop}%
\bibitem [{\citenamefont {Hormoz}\ and\ \citenamefont
  {Brenner}(2011)}]{hormoz2011design}%
  \BibitemOpen
  \bibfield  {author} {\bibinfo {author} {\bibfnamefont {S.}~\bibnamefont
  {Hormoz}}\ and\ \bibinfo {author} {\bibfnamefont {M.~P.}\ \bibnamefont
  {Brenner}},\ }\href@noop {} {\bibfield  {journal} {\bibinfo  {journal} {Proc.
  Natl. Acad. Sci. U.S.A.}\ }\textbf {\bibinfo {volume} {108}},\ \bibinfo
  {pages} {5193} (\bibinfo {year} {2011})}\BibitemShut {NoStop}%
\bibitem [{\citenamefont {Hedges}, \citenamefont {Mannige},\ and\ \citenamefont
  {Whitelam}(2014)}]{hedges2014growth}%
  \BibitemOpen
  \bibfield  {author} {\bibinfo {author} {\bibfnamefont {L.~O.}\ \bibnamefont
  {Hedges}}, \bibinfo {author} {\bibfnamefont {R.~V.}\ \bibnamefont {Mannige}},
  \ and\ \bibinfo {author} {\bibfnamefont {S.}~\bibnamefont {Whitelam}},\
  }\href@noop {} {\bibfield  {journal} {\bibinfo  {journal} {Soft Matter}\
  }\textbf {\bibinfo {volume} {10}},\ \bibinfo {pages} {6404} (\bibinfo {year}
  {2014})}\BibitemShut {NoStop}%
\bibitem [{\citenamefont {Einstein}(1905)}]{einstein1905molekularkinetischen}%
  \BibitemOpen
  \bibfield  {author} {\bibinfo {author} {\bibfnamefont {A.}~\bibnamefont
  {Einstein}},\ }\href@noop {} {\bibfield  {journal} {\bibinfo  {journal} {Ann.
  Phys.}\ }\textbf {\bibinfo {volume} {322}},\ \bibinfo {pages} {549} (\bibinfo
  {year} {1905})}\BibitemShut {NoStop}%
\bibitem [{\citenamefont {Li}\ \emph {et~al.}(2014)\citenamefont {Li},
  \citenamefont {Yang}, \citenamefont {Jiang}, \citenamefont {Yan},\ and\
  \citenamefont {Liu}}]{li2014controlled}%
  \BibitemOpen
  \bibfield  {author} {\bibinfo {author} {\bibfnamefont {W.}~\bibnamefont
  {Li}}, \bibinfo {author} {\bibfnamefont {Y.}~\bibnamefont {Yang}}, \bibinfo
  {author} {\bibfnamefont {S.}~\bibnamefont {Jiang}}, \bibinfo {author}
  {\bibfnamefont {H.}~\bibnamefont {Yan}}, \ and\ \bibinfo {author}
  {\bibfnamefont {Y.}~\bibnamefont {Liu}},\ }\href@noop {} {\bibfield
  {journal} {\bibinfo  {journal} {J. Am. Chem. Soc.}\ }\textbf {\bibinfo
  {volume} {136}},\ \bibinfo {pages} {3724} (\bibinfo {year}
  {2014})}\BibitemShut {NoStop}%
\bibitem [{\citenamefont {SantaLucia~Jr}\ and\ \citenamefont
  {Hicks}(2004)}]{santalucia2004thermodynamics}%
  \BibitemOpen
  \bibfield  {author} {\bibinfo {author} {\bibfnamefont {J.}~\bibnamefont
  {SantaLucia~Jr}}\ and\ \bibinfo {author} {\bibfnamefont {D.}~\bibnamefont
  {Hicks}},\ }\href@noop {} {\bibfield  {journal} {\bibinfo  {journal} {Annu.
  Rev. Biophys. Biomol. Struct.}\ }\textbf {\bibinfo {volume} {33}},\ \bibinfo
  {pages} {415} (\bibinfo {year} {2004})}\BibitemShut {NoStop}%
\bibitem [{\citenamefont {Koehler}\ and\ \citenamefont
  {Peyret}(2005)}]{koehler2005thermodynamic}%
  \BibitemOpen
  \bibfield  {author} {\bibinfo {author} {\bibfnamefont {R.~T.}\ \bibnamefont
  {Koehler}}\ and\ \bibinfo {author} {\bibfnamefont {N.}~\bibnamefont
  {Peyret}},\ }\href@noop {} {\bibfield  {journal} {\bibinfo  {journal}
  {Bioinformatics}\ }\textbf {\bibinfo {volume} {21}},\ \bibinfo {pages} {3333}
  (\bibinfo {year} {2005})}\BibitemShut {NoStop}%
\bibitem [{\citenamefont {Sobczak}\ \emph {et~al.}(2012)\citenamefont
  {Sobczak}, \citenamefont {Martin}, \citenamefont {Gerling},\ and\
  \citenamefont {Dietz}}]{sobczak2012rapid}%
  \BibitemOpen
  \bibfield  {author} {\bibinfo {author} {\bibfnamefont {J.-P.~J.}\
  \bibnamefont {Sobczak}}, \bibinfo {author} {\bibfnamefont {T.~G.}\
  \bibnamefont {Martin}}, \bibinfo {author} {\bibfnamefont {T.}~\bibnamefont
  {Gerling}}, \ and\ \bibinfo {author} {\bibfnamefont {H.}~\bibnamefont
  {Dietz}},\ }\href@noop {} {\bibfield  {journal} {\bibinfo  {journal}
  {Science}\ }\textbf {\bibinfo {volume} {338}},\ \bibinfo {pages} {1458(SI)}
  (\bibinfo {year} {2012})}\BibitemShut {NoStop}%
\bibitem [{\citenamefont {Shneidman}(2003)}]{shneidman2003lowest}%
  \BibitemOpen
  \bibfield  {author} {\bibinfo {author} {\bibfnamefont {V.~A.}\ \bibnamefont
  {Shneidman}},\ }\href@noop {} {\bibfield  {journal} {\bibinfo  {journal} {J.
  Stat. Phys.}\ }\textbf {\bibinfo {volume} {112}},\ \bibinfo {pages} {293}
  (\bibinfo {year} {2003})}\BibitemShut {NoStop}%
\bibitem [{\citenamefont {Bondy}\ and\ \citenamefont
  {Murty}(1976)}]{bondy1976graph}%
  \BibitemOpen
  \bibfield  {author} {\bibinfo {author} {\bibfnamefont {J.~A.}\ \bibnamefont
  {Bondy}}\ and\ \bibinfo {author} {\bibfnamefont {U.~S.~R.}\ \bibnamefont
  {Murty}},\ }\href@noop {} {\emph {\bibinfo {title} {Graph Theory with
  Applications}}},\ Vol.~\bibinfo {volume} {6}\ (\bibinfo  {publisher}
  {Macmillan, London},\ \bibinfo {year} {1976})\BibitemShut {NoStop}%
\end{thebibliography}
\end{document}